\documentclass[11pt,a4paper]{article}
\usepackage{jheppub}

\usepackage{graphicx,bm}
\usepackage{amsmath,amssymb,mathrsfs,verbatim,subfigure}

\title{Conductivity and quasinormal modes in holographic theories}

\author{M.A.~Stephanov,}
\author{Y.~Yin}

\affiliation{Department of Physics, University of Illinois,
Chicago, IL 60607-7059, USA}

\emailAdd{misha@uic.edu}
\emailAdd{yyin3@uic.edu}

\abstract{
  We show that in field theories with a holographic dual the retarded
  Green's function of a conserved current can be represented as a convergent sum
  over the quasinormal modes. We find that the zero-frequency
  conductivity is related to the sum over quasinormal modes and their
  high-frequency asymptotics via a sum rule. We derive the asymptotics
  of the quasinormal mode frequencies and their residues using the
   phase-integral (WKB) approach and provide analytic insight
  into the existing numerical observations concerning the asymptotic
  behavior of the spectral densities.
}


\arxivnumber{}

\begin{document}
\maketitle

\section{Introduction and summary of results}

The transport properties of the strongly coupled quark-gluon plasma (sQGP) created at
RHIC~\cite{Adcox:2004mh,Back:2004je,Arsene:2004fa,Adams:2005dq}
attracted much attention recently. One of the most important transport
parameters is the conductivity $ \sigma $ associated with a conserved
vector current. For example, the quark current conductivity is an
indicator of deconfinement. Furthermore, the conductivity of
the current of light quarks can be related, via Kubo formula, to the
soft limit of the thermal photon production rate by QCD plasma. In addition,
the Einstein relation equates conductivity to the product of quark
susceptibility and quark diffusion constant. For heavy quarks, the
diffusion constant is an important quantity characterizing medium
effects on quark propagation. The results of application of various
phenomenological models to that problem is best expressed in terms of
the diffusion constant~\cite{Petreczky:2005nh}.

Due to strong coupling, the calculation of conductivity in QCD at
temperatures relevant to the experiments is a challenging
task. Lattice calculations, being restricted to the finite interval of
Euclidean time, require analytic continuation to infinitely large real
time in order to determine transport coefficients such as
conductivity. Some interesting results have been obtained assuming
certain analytic behavior~\cite{Aarts:2007wj}. Clearly, it would be
greatly helpful to better understand the analytic properties of the
current-current correlator and find generic, model-independent constraints such as sum rules on the conductivity.

As a step towards this goal, in this paper, we consider the class of
quantum field theories in $3+1$ space-time dimensions whose
correlation functions can be computed via AdS/CFT duality
~\cite{Maldacena:1997re, Gubser:1998bc, Witten:1998qj}. Such theories
have been employed to describe thermodynamics and transport in
strongly coupled regime of QCD. We consider a generic gravity dual
set-up satisfying some mild technical assumptions as considered in
Ref.~\cite{Gulotta:2010cu}. One can show that the retarded Green's
function at vanishing momentum $ G_R(\omega) $ of a conserved vector
current $J^{\mu}$ calculated from such gravity background is a
meromorphic function with infinite number of simple poles located in
the lower half-plane~\cite{Gulotta:2010cu}. In the context of
gauge/gravity correspondence, those poles are referred to as quasi-normal 
modes~\cite{Berti:2009kk,Horowitz:1999jd, Nunez:2003eq,
Kovtun:2005ev, Starinets:2002br,Birmingham:2001pj}.

We show that the Green's function $G_R(\omega)$ can be represented as a
{\em convergent} sum over its poles:
\begin{equation}\label{eq:grep} 
G_R(\omega)=-i\sigma\omega+ C\omega^2 + \omega^3\sum_{n}\left[\frac{
    r_n}{\omega_n\left(\omega-\omega_n\right)}+\frac{\tilde{r}_n}{\tilde{\omega}_n\left(\omega-\tilde{\omega}_n\right)}\right]
.
\end{equation}
in terms of conductivity $\sigma$ and the residues $r_n$. The real
coefficient $C$   depends on the definition of $G_R(\omega)$, however,
its temperature-dependent part is fixed by parameters of the second
order hydrodyanmics~\cite{Baier:2007ix} (and equals $\sigma$ times $\tau_j$ defined in Ref.~\cite{Hong:2010at}).
Since $ G_R(\omega)$ has a ``mirror'' symmetry: $
G^{*}_R(\omega)=G_R(\tilde{\omega})$ where $ \tilde{\omega}\equiv
-\omega^{*} $, if $ \omega_n $ is a pole of $ G_R(\omega)$, so is $
\tilde{\omega}_n $.\footnote{For simplicity, we assume there is no
  pole located at the negative imaginary axis thus $\omega_n\neq
  \tilde{\omega}_n$. If there is, the modification of our treatment
  here is trivial.}  The index $n$ in the infinite sum in
Eq.~(\ref{eq:grep}) counts  poles located in the fourth quadrant of
the complex $ \omega $ plane. The contribution of the mirror poles is
the second term in the infinite sum. 

Expansion of $ G_R(\omega) $ in terms of poles corresponding to quasinormal modes has been
suggested in Ref.~\cite{Teaney:2006nc, Amado:2007yr}. However, to
avoid ambiguities, one needs to show that the summation in the
expansion is convergent for any finite $ \omega $ away from the
poles. To this end, we
have established the convergence by determining the large $n$ asymptotics
of both $\omega_n$ and $r_n$ (by extending previous
work~\cite{Natario:2004jd} on asymptotics of $ \omega_n$):
\begin{equation}\label{eq:gapoff}
\omega_n \to n\omega_0 +\Delta, \qquad r_n\to K\omega_0,\qquad \text{when}\qquad n\to\infty.
\end{equation} 
The complex numbers $ \omega_0 $ and $ \Delta $ are sometimes called (asymptotic) ``gap'' and ``offset'' of the quasi-normal modes respectively
~\cite{Natario:2004jd}. The coefficient $K$ is related to the leading asymptotic
behavior of $G_R$, which in the deep Euclidean regime
$\omega\to i\infty$ is given by the operator product expansion (OPE): 
\begin{equation}\label{eq:OPE0}
{G}_{R}(i\omega_E)\to-{2K}\omega_E^2\log\omega_E 
,\qquad \omega_E\to\infty.
\end{equation}
The constant $K$ is proportional to the number of the charge
carrying degrees of freedom.

Finally, by matching the asymptotic behavior in the deep Euclidean
regime of the representation~(\ref{eq:grep}) to the OPE~(\ref{eq:OPE0})
we derive a relationship between the conductivity and the quasinormal
modes:
\begin{equation}\label{eq:srs}
\sigma =-K{\rm Im\,}\left(\omega_0 +2\Delta\right)+2\sum_{n}{\rm Im\,}\left(r_n-K\omega_0\right).
\end{equation}
This paper is organized as follows. Sec.~\ref{sec:derivation}
presents  the derivation of the representation and the sum rule. In
Sec.~\ref{sec:asymptotic}, we establish the asymptotics of $
\omega_n,r_n $ using the WKB approximation. In
Sec.~\ref{sec:softwall}, we investigate how the sum rule is
saturated by studying the ``soft-wall" model~\cite{Karch:2006pv} at
finite temperature numerically. We summarize and explain qualitatively
and quantitatively how the asymptotic behavior of $\omega_n,r_n$ is
related to the ``damped oscillating" behavior~\cite{Teaney:2006nc} of
spectral densities in Sec.~\ref{sec:conclusion}. In
Appendix~\ref{sec:redundant}, we clarify a subtle point in the
holographic calculation of the retarded correlators in the lower half
of the complex $\omega$ plane. In Appendix~\ref{sec:stokes} we
derive the Stokes constant formula we used in the WKB calculation. We also formulate a family of f-sum rules from holography in Appendix~\ref{sec:fsum}.  
  
\section{The derivation of the representation
  and the sum rule.}
\label{sec:derivation}

\subsection{The representation}
\label{sec:der-rep}

We study the retarded Green's function $ G_R(\omega) $ of a spatial conserved vector current operator $ J^{1}$ at zero three-momentum and the corresponding spectral function $\rho(\omega) $:
\begin{equation}
G_R(\omega)=-i\int dt\, e^{i\omega t}\,\theta(t)\langle
[J^{1}(t),J^{1}(0)]\rangle,\qquad \rho(\omega)=-{\rm Im\,} G_R(\omega).
\end{equation}
We assume that the quantum field theory under consideration has a
holographically dual description. As discussed in
Ref.~\cite{Gulotta:2010cu},  $G_R(\omega)$ calculated from holography,
is a meromorphic function on general grounds. We could thus consider
the following Mittag-Leffler expansion of $G_R(\omega) $ modulo
contact terms:
\begin{equation}\label{eq:gsum}
\bar G_R\equiv \frac{G_R(\omega)}{\omega^2}
=-i\,\frac{\sigma}{\omega}
+ \omega\sum_{n}\left[\frac{
    r_n}{\omega_n\left(\omega-\omega_n\right)}+\frac{\tilde{r}_n}{\tilde{\omega}_n\left(\omega-\tilde{\omega}_n\right)}\right]
+P(\omega),
\end{equation}
where $P(\omega)$ is a polynomial of $\omega$. The scaled retarded
Green's function $ \bar{G}_R(\omega)$ defined in Eq.~(\ref{eq:gsum})
has a pole at $\omega=0$ with residue related to the conductivity by
the usual Kubo formula:
\begin{equation}\label{eq:conductivity}
\sigma=\mathop{\lim}\limits_{\omega \to 0}\frac{\rho(\omega)}{\omega}
\end{equation}
while the quasinormal mode residues are defined as 
\begin{equation}\label{eq:rn}
r_n=\mathop{\lim}\limits_{\omega \to \omega_n}\left(\omega-\omega_n\right)\bar{G}_R(\omega).
\end{equation}
In the deep Euclidean region, in accordance with the operator product
expansion (OPE), $ G_R(\omega) $ has the following asymptotics:
\begin{equation}\label{eq:OPE}
\bar{G}_{R}(i\omega_E)={2K}\log\omega_E +{\rm const}+O(\omega_E^{-2}),\qquad \omega_E\to\infty
\end{equation}
where the leading contribution is from the unit operator. Here we have used the relation
$G_R(i\omega_E)=-G_E(\omega_E>0)$ where $G_E(\omega_E)$ is the
Euclidean correlator \footnote{We analytically continue the Euclidean
  correlator $G_E(\omega_E)$ from the discrete set of Matsubara
  frequencies.} and the OPE of $G_E(\omega_E)$. When writing down Eq.~(\ref{eq:OPE}), we have assumed that the lowest dimension of those non-trivial operators entering the OPE of
$G_E(\omega_E)$ is no less than $2$. That fact is crucial for subsequent discussions. 

Since $ P(\omega)$ is a polynomial, the logarithmic behavior in
Eq.~(\ref{eq:OPE}) should be matched by the summation over
pole contributions in the representation ~(\ref{eq:gsum}). Thus the
number of poles has to be infinite. As we will show in the next
section, $\omega_n, r_n$ have the asymptotic behavior given by
Eq.~(\ref{eq:gapoff}).
As a result, for any finite $ \omega$ away from
$\omega_n(\tilde{\omega}_n)$, the summation of $
r_n/\omega_n(\omega-\omega_n)$ in Eq.~(\ref{eq:gsum}) is convergent. 

We also note from Eq.~(\ref{eq:OPE}) that the polynomial $ P(\omega) $ cannot grow faster than a constant. Consequently, it should be a real constant, i.e. $C$, as required by the ``mirror" symmetry of $ G_R(\omega)$. We thus establish the representation~(\ref{eq:grep}).

\subsection{The sum rule}
\label{sec:der-sum-rule}

In order to derive the sum rule relating conductivity $\sigma$ to the
quasinormal modes, we shall match the asymptotic behavior of
representation in Eq.~(\ref{eq:gsum}) to the OPE
Eq.~(\ref{eq:OPE}).

To facilitate the matching, we apply  ``Borel'' transformation~\cite{Shifman:1978bx} defined by:
\begin{equation}
\hat{\mathcal{B}}_{1/t_B}=\frac{\omega_E^{n}}{(n-1)!}\left(-\frac{d}{d\omega_E}\right)^n,\qquad
\text{when}\qquad \omega_E\to +\infty,\ n\to\infty,\ \frac{\omega_E}{n}=1/t_B,
\end{equation}
to $\bar{G}_{R}(\omega)$ in the deep Euclidean region:
\begin{equation}\label{eq:Zt}
\hat{\mathcal{B}}_{1/t_B}\bar{G}_R(i\omega_E)= -t_B\sigma
-it_B\sum_{n}\left(r_n e^{-i\omega_n t_B}+\tilde{r}_n
e^{-i\tilde{\omega}_nt_B}\right)
=-t_B\sigma+2t_B\sum_{n}{\rm Im\,}\left(r_n e^{-i\omega_n t_B}\right).
\end{equation}
All relevant formulas for the Borel transformation are listed in Appendix.~\ref{sec:fsum}. 
For any positive $ t_B $, the sum in Eq.~(\ref{eq:Zt}) is convergent
since $ {\rm Im\,}\omega_n<0 $. Applying the Borel transformation to the asymptotic expansion~(\ref{eq:OPE}), we obtain for small $ t_B $ :
\begin{equation}\label{eq:BG-OPE}
\hat{\mathcal{B}}_{1/t_B}\bar{G}_R(i\omega_E)=-{2K}+O(t_B^2),\qquad\text{when}\qquad t_B\to 0^{+}.
\end{equation}
Matching Eq.~(\ref{eq:Zt}) and Eq.~(\ref{eq:BG-OPE}) at small $t_B$, we find:
\begin{equation}\label{eq:sumr1}
\sigma =2\mathop{\lim}\limits_{t_B \to 0^{+}}\left[\,{\rm Im\,}\sum_{n}r_n e^{-i\omega_n t_B}+\frac{K}{t_B}\right].
\end{equation}
One can check, using  Eq.~(\ref{eq:gapoff}),  that when $t_B \to
0^{+}$, the $1/t_B$ divergence in Eq.~(\ref{eq:sumr1}) is canceled as
it should be.

Using the asymptotic behavior of $r_n$ in Eq.~(\ref{eq:gapoff}) we can
evaluate R.H.S of Eq.~(\ref{eq:sumr1}) by rearranging the infinite sum as
\begin{equation}\label{eq:Im-sum-rn}
\sigma =2\mathop{\lim}\limits_{t_B \to 0^{+}}\left[{\rm Im\,}\sum_{n}\left(r_n-K\omega_0\right) e^{-i\omega_n t_B}
+K\,{\rm Im\,}\sum_{n} \omega_0 e^{-i\omega_n t_B}
+\frac{K}{t_B}\right].
\end{equation}
The summation of $ (r_n-K\omega_0) $ is convergent due to
Eq.~(\ref{eq:gapoff}) (see discussion in
Sec.~(\ref{sec:asymptotic})). Consequently one can exchange the
sequence of summation and taking $ t_B\to 0^{+} $ limit. The second sum in Eq.~(\ref{eq:Im-sum-rn}) can be evaluated
explicitly for  $t_B\to0^+$. Its divergence $-K/t_B$ is cancelled by
the last term in Eq.~(\ref{eq:sumr1}) and the remaining
finite part can be obtained using asymptotics of $\omega_n$:
\begin{multline}
\mathop{\lim}\limits_{t_B \to 0^{+}}\left[\,{\rm Im\,}\sum_{n}
  {\omega_0}e^{-i\omega_n t_B}+\frac{1}{t_B}\right]
=\mathop{\lim}\limits_{t_B \to 0^{+}}\left[\,{\rm
    Im\,}\sum^{\infty}_{n=1} {\omega_0} e^{-i(n\omega_0+\Delta)
    t_B}+\frac{1}{t_B}\right]\\
=\mathop{\lim}\limits_{t_B \to
  0^{+}}\left[\,{\rm
    Im\,}\frac{\omega_0e^{-i\Delta t_B}}{e^{i\omega_0 t_B}-1}+\frac{1}{t_B}\right]
\end{multline}
Expanding the expression in brakets around $ t_B=0 $\footnote{As a
  side, we note that the radius of convergence of the Taylor expansion
  in $t_B$ is $ |2\pi/\omega_0| $ because of a pole at $
  t_B=2\pi/\omega_0 $. That suggests that $ |\omega_0|/2\pi $ sets a
  scale below which the asymptotic expansion (or OPE) of $
  G_R(\omega_E) $ will be broken.}, taking the limit and substituting
into Eq.~(\ref{eq:Im-sum-rn}), we obtain the sum rule Eq.~(\ref{eq:srs}).

\subsection{$\mathcal{N}=4$ SYM theory in large $ N_c $, strongly coupling limit as an example }
In Ref.~\cite{Myers2007}, $G_R(\omega)$ in $\mathcal{N}=4$ SYM theory
is derived in large $ N_c $, strong coupling limit using AdS/CFT correspondence:
\begin{equation}\label{eq:adsg}
G_{R}(\omega)=\frac{N^2_cT^2}{8}\left\{\frac{i\omega}{2\pi T} +\frac{\omega^2}{(2\pi T)^2}\left[\psi\left(\frac{(1-i)\omega}{4\pi T}\right)+\psi\left(-\frac{(1+i)\omega}{4\pi T}\right) \right]\right\}
\end{equation}
with  $ \psi $ the logarithmic derivative of the gamma function. This retarded correlator has a quasi-normal spectrum with:
\begin{equation}\label{eq:qnma}
\omega_n= 2\pi T(1-i)n,\qquad r_n= \frac{N^2_c}{16\pi^2}(1-i)T.
\end{equation}
Therefore, for this theory, $ \omega_0=2(1-i)\pi T, \Delta=0 $ and $K=N^2_c/32\pi^2$. From Eq.~(\ref{eq:adsg}), we also have $\sigma=N^2_cT/16\pi$. One sees immediately that sum rule~(\ref{eq:srs}) holds.

\section{\label{sec:asymptotic}The asymptotics of quasinormal
  frequencies and residues}
\subsection{The Green function in the holographically dual description}
To calculate $ G_R(\omega) $ using gauge-gravity/holographic
correspondence we need to consider the second order variation of the
5-dimensional bulk action with respect to the bulk gauge field $ V_M $
dual to the vector current $ J^{\mu} $ in the boundary theory. The
relevant part of the bulk action has the usual Maxwell form:
\begin{equation}\label{eq:action}
S=-\frac{1}{4g^2_5}\int d^5x\sqrt{g}e^{-\phi}V_{MN}V^{MN}
\end{equation}
where $ g^2_5 $ is the 5D gauge coupling, $\phi$ is the background scalar
field which, in general, is a combination of dilaton and/or tachyon
fields, corresponding to the conformal and/or chiral symmetry breaking, and $ V_{MN}=\partial_MV_N-\partial_NV_M$. We consider the most general metric (up to general coordinate transformations) possessing three-dimensional (3D) Euclidean isometry:
\begin{equation}\label{eq:gmetric}
ds^2=e^{2A(z)}\left(hdt^2-d\vec{x}^2-h^{-1}dz^2\right).
\end{equation}
The equation of motion resulting from the action~(\ref{eq:action}) reads:
\begin{equation}\label{eq:EOM}
\partial_{z}(he^{B}\partial_z V)+\omega^2 h^{-1}e^{B}V=0
\end{equation} 
where $ V=V_1$ and $B=A-\phi$. As usual, the thermal bath is
represented by the black brane, corresponding to a real positive zero
of $h(z)$, with  temperature $ T $ given by:\footnote{For simplicity, we assume that $ h(z) $ has only one real positive zero.}
\begin{equation}\label{eq:Tem}
T=\frac{1}{4\pi}|h'(z_H)|.
\end{equation}
Here and hereafter a prime denotes the derivative with respect to $ z
$.  Also in this section, we will set $\pi T=1$ for convenience, i.e.,
we will measure all dimensional quantities in units of $\pi T$. We
require the background to be AdS in the asymptotic at the boundary:
\begin{equation}\label{eq:B}
B(z)=-\log z,\qquad\text{when}\qquad z\to 0.
\end{equation}
Then $z=0$ and $ z=z_H$ are two regular singular points of Eq.~(\ref{eq:EOM}).
The retarded Green's function, up to a contact term, is given by the standard holographic prescription ~\cite{Son:2002sd, Herzog:2002pc}:
\begin{equation}\label{eq:green}
G_R(\omega)=-\frac{1}{g^2_5}\mathop{\lim}\limits_{z \to 0}\left[he^{B}\frac{V_{-}'(z,\omega)}{V_{-}(z,\omega)}+\omega^2\log z\right]
\end{equation}
where $ V_{-} $ denotes the Frobenius power series solution near $ z_H
$ of indicial exponents $-i\omega/4\pi T$, i.e., $ V_{-}\sim
(z-z_H)^{-i\omega/4\pi T} [1+O(z-z_H)]$, corresponding to an in-falling
wave. Further assuming the Frobenius power series solutions at $z=0$ and $z=z_H$ have an overlapping region of validity along the real axis $0<z<z_H$, one can show, along the lines of Ref.~\cite{Gulotta:2010cu}, that $G_R(\omega)$ is a meromorphic function\footnote{That conclusion from Ref.~\cite{Gulotta:2010cu} has some uncertainties at a set of discrete frequencies $\omega=-2in\pi T$ where the difference between two indicial exponents $r_+-r_{-}$ is an integer. We will settle that subtle issue in Appendix.~\ref{sec:redundant}}.

\subsection{Near-boundary asymptotics}
\label{sec:asymptotics}

To establish the large $n$ asymptotics of quasinormal mode parameters in Eq.~(\ref{eq:gapoff}) we first introduce the book-keeping Schr{\"o}dinger coordinate
\begin{equation}\label{eq:xi}
\xi(z)=\int^{z}_0dz'\frac{1}{h(z')}
\end{equation}
and a wave-function-like 
\begin{equation}\label{eq:bigpsi}
\Psi(\xi)=e^{B(z)/2}V(z).
\end{equation}
Then Eq.~(\ref{eq:EOM}) is brought into the standard Schr{\"o}dinger-like form:
\begin{equation}\label{eq:schrodinger}
\frac{d^2\Psi(\xi)}{d\xi^2}+\left(\omega^2 -U(\xi)\right)\Psi(\xi)=0
\end{equation}
where the potential $ U(z) $, as a function of $ z $, is given by
\begin{equation}\label{eq:schpotential}
U(z)=h^2\left[\left(\frac{B'}{2}\right)^2+\frac{B''}{2}+\frac{h'B'}{2h}\right].
\end{equation}
Near the boundary $ z=\xi=0 $, Eq.~(\ref{eq:schrodinger}) becomes
\begin{equation}\label{eq:near0eq}
\frac{d^2\Psi(\xi)}{d\xi^2}+\left(\omega^2 -\frac{\nu^2_{0}-\frac{1}{4}}{\xi^2}\right)\Psi(\xi)=0,
\end{equation}
where $\nu_{0}=1$ (the same as the spin of the fluctuations we are
studying).\footnote{Our formalism below can be readily generalized to
  other values of $\nu_{0}$.} Its solutions are:
\begin{equation}\label{eq:Bessels}
\Psi(\xi)=A_{+}(\omega)\sqrt{\frac{\pi\omega \xi}{2}}H^{(1)}_{\nu_0}(\omega \xi)+A_{-}(\omega)\sqrt{\frac{\pi\omega \xi}{2}}H^{(2)}_{\nu_0}(\omega \xi)
\end{equation}
with $ H^{(1)}_\nu,H^{(2)}_\nu $ denoting the Hankel functions of the first and second kind respectively. Using the definition~(\ref{eq:green}), we can calculate the asymptotic behavior of $\bar{G}_{R}(\omega) $ from the solution~(\ref{eq:Bessels}):
\begin{equation}\label{eq:wkbg}
\bar{G}_{R}(\omega)={2K}\left[\gamma+\log(\omega /2)+\frac{\pi }{2\tan[\omega\mathcal{D}(\omega)]}\right]+O(\omega^{-2}),\qquad\text{with}\qquad K=\frac{1}{2g^2_5}.
\end{equation}
Here, $ \gamma $ is the Euler-Mascheroni
constant. 
The function $ \mathcal{D}(\omega) $ is defined by
\begin{equation}\label{eq:D}
e^{-2i\omega\mathcal{D}(\omega)}\equiv\frac{A_{+}(\omega)}{A_{-}(\omega)}.
\end{equation}
It will be determined by applying the in-falling wave boundary
condition near the horizon. The poles of $ \bar G_R(\omega) $, $\omega_n$,
as well as the residues $ r_n$, for sufficiently large $n$ can be determined from Eq.~(\ref{eq:wkbg}):
\begin{equation}\label{eq:qnmd}
\omega_n\mathcal{D}(\omega_n)=n\pi, \qquad r_n=\frac{{\pi} K}{\partial_{\omega}(\omega\mathcal{D}(\omega))|_{\omega=\omega_n}}.
\end{equation}
One may note that corrections in Eq.~(\ref{eq:wkbg}) are $O(\omega^{-2})$
as we are required to match Eq.~(\ref{eq:wkbg}) with the OPE results
Eq.~(\ref{eq:OPE})\footnote{From the gravity side, that condition will
  be satisfied if $U(\xi)-(\nu^2_0-1/4)/\xi^2$ is bounded near the boundary.}. As a result, $\omega_n, r_n$ calculated from Eq.~(\ref{eq:qnmd}) are accurate up to (including) the order of $n^{-1}$ relative to the corresponding leading large $n$ results.
 
To determine $\omega_n$ and $r_n$ from Eq.~(\ref{eq:qnmd}) for large $n$, we need to know asymptotics of $ \mathcal{D}(\omega)$. We
shall determine it by following the solution of the Schr{\"o}dinger
equation along a path from $z=0$ to $z=z_H$ where we apply the
in-falling wave boundary condition. We shall use the WKB solution
along that path. Thus it is important that the region, which we denote
$ R_0 $, where $|z|\ll 1 $, does overlap with the region $R_1$,
defined by $ |\omega^2|\gg |U(\xi)| $, where WKB approximation is
applicable, for sufficiently large $\omega$.
We denote that overlapping region by $ R_2 $. In $R_2$ $
|\omega\xi|\gg 1 $. (The above definitions of the regions are
summarized in Table.~\ref{region}). 

Finally, due to the asymptotic behavior of
the Hankel functions at large argument,
\begin{equation}\label{eq:Bessela}
H^{(1)}_\nu(x)\approx \sqrt{\frac{2}{\pi x}}\exp(ix-i\delta),
\qquad H^{(2)}_\nu\approx\sqrt{\frac{2}{\pi x}}\exp(-ix+i\delta)
\qquad (|x|\gg1),
\end{equation}
where  $ \delta=(2\nu +1)\pi/4$, we find for large $|\omega\xi|$
(i.e., in the region $R_2$):
\begin{equation}\label{eq:psia}
\Psi(\xi)=A_{+}e^{i\omega\xi-i\delta_0}+A_{-}e^{-i\omega\xi+i\delta_0},\qquad\text{with}\qquad\xi\in R_2
\end{equation}
where $\delta_0=3\pi/4$. We shall match the WKB solution to this asymptotics.

\begin{table}
\centering
\caption{\label{region} The definition of different regions.}
\begin{tabular}{|l|l|}
\hline
$R_0$& $|z|\ll1$ and Eq.~(\ref{eq:schrodinger}) is well approximated by Eq.~(\ref{eq:Bessela}).\\
\hline
$R_1$& $|\omega^2|\gg |U(\xi)|$ and one can use the WKB approximation.\\
\hline
$R_2$& $R_0\cap R_1$ where we will match Eq.~(\ref{eq:WKB1}) and Eq.~(\ref{eq:psia}).\\
\hline
\end{tabular}
\end{table}

\subsection{The WKB approximation} 	

In region $ R_1 $ one can use WKB approximation
(also known as the phase integral method~\cite{headingbook})  to solve
Eq.~(\ref{eq:schrodinger}). The application of the method to calculating the
asymptotic quasi-normal modes is reviewed in
Ref.~\cite{Natario:2004jd}. 

The two linearly independent WKB solutions are
given by $Q^{-1/2}\exp[\pm i\int\! d\xi\, Q]$, where 
\begin{equation}\label{eq:Q-def}
Q^2(\xi)=\omega^2 -U(\xi)
\end{equation} 
These solutions are singular at points where $Q=0$ -- the {\em turning
  points}.  At a turning point the WKB approximation breaks down.  We
shall assume that a generic Schr{\"o}dinger potential grows as
$U(z)\sim z^m $ when $ |z| $ is large, where $ m $ is a positive real
number. Then in the limit of large $ |\omega| $, there will be turning
points determined by the condition $ \omega^2-z^m=0$, which has
multiple solutions. We denote one such turning point by $z_T$ and map
it to the Schr{\"o}dinger coordinate $\xi_T=\xi(z_T)$.

\begin{figure}[htb]
  \centering 
	\subfigure[]{\label{fig:inz}
	\includegraphics[width=.45\textwidth]{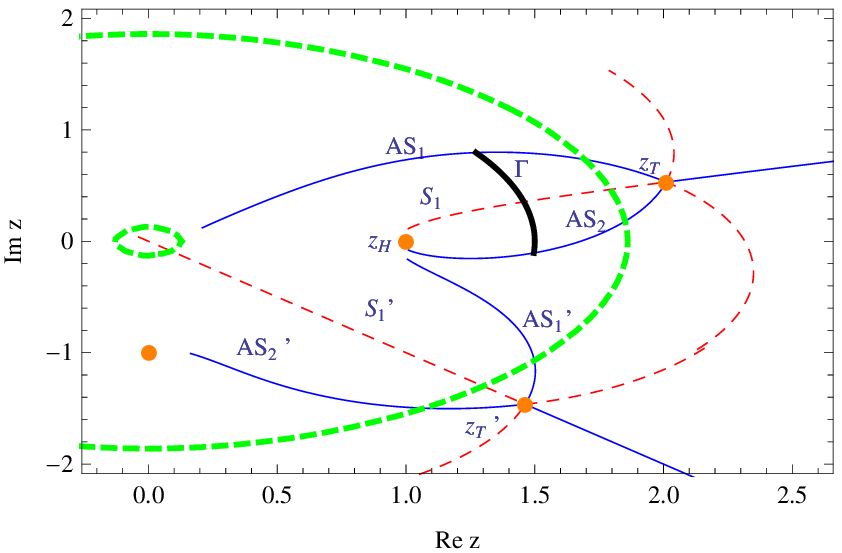}}
	\subfigure[]{\label{fig:inxi}
    \includegraphics[width=.45\textwidth]{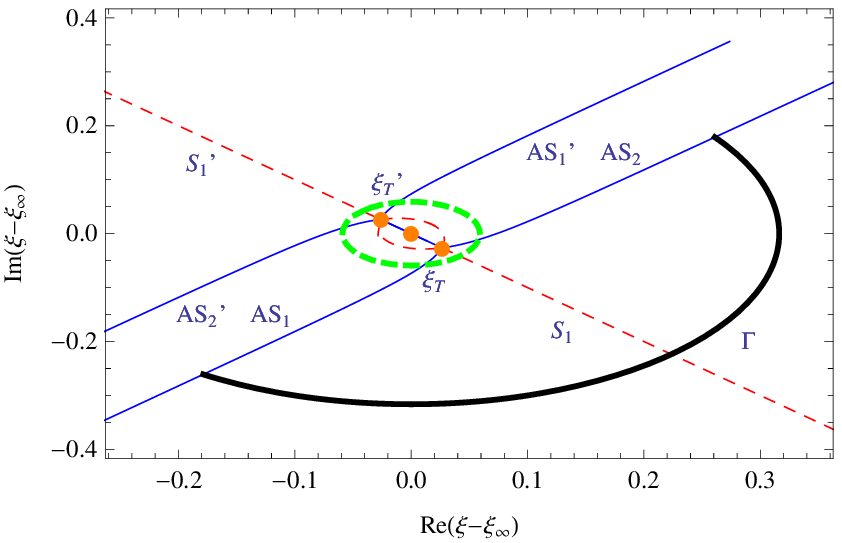}}
\caption{ 
\label{fig:stokes} 
The Stokes diagram for $\omega=10(1-i)\pi T$ of the pure AdS black-hole background in the complex $z$ plane (left panel) and in the complex $\xi$ plane (right panel). We set $\pi T=1$.  Anti-Stokes lines $ AS_1,AS_2,AS_3 $ are blue solid lines while Stokes lines $ S_1,S_2,S_3 $ are red dashed lines. The path $\Gamma$ connecting $AS_1$ and $AS_2$ is plotted as a thick black line. To determine the asymptotics~(\ref{eq:gapoff}), we connect the region $R_2$ (around $z=0$) and horizon $z_H$ via the path along $AS_1, \Gamma$ and $AS_2$. We sketch the boundary of $R_1$ (where the WKB approximation applies) schematically using the green thick dashed lines. For completeness, we plot Stokes lines and anti-Stokes lines emanating from another turning point $z'_T$ next to $z_T$.
}
\end{figure}

The exponential
in the WKB solutions is purely oscillatory along an integral curve
defined by condition
\begin{equation}
{\rm Im\,}\int^{\xi}_{\xi_T} d\xi' Q(\xi')=0
\end{equation}
This condition defines the anti-Stokes line(s), $ AS$, with respect to the point $
\xi_T(z_T) $. For a simple zero $ \xi_T(z_T)$ of $ Q(\xi(z))$, there
will be three anti-Stokes lines $ AS_1,AS_2,AS_3 $ emanating from
it. We choose $z_T$ (or $\xi_T$) among solutions of $Q(\xi(z))=0$ by
requiring that $ AS_{1} $ has an overlap with $ R_2 $ while $ AS_{2} $
ends on $ z_H$ as illustrated by Fig.~\ref{fig:inz}. As it will
become clear soon, for large $ |\omega| $, the existence of such a
turning point $ z_T$ (or $\xi_T$) is necessary to solve the condition~(\ref{eq:qnmd}).

We shall define point $\xi_\infty$ as the limit
\begin{equation}\label{eq:xiinf}
\xi_\infty\equiv\mathop{\lim}\limits_{|\omega| \to \infty}\xi_T =\mathop{\lim}\limits_{|\omega| \to \infty}\int^{z_T}_{0}\frac{dz}{h(z)}
\end{equation}
which exists if the integration in the above equation is convergent. 
To avoid ambiguity, we specify the path of the integration in Eq.~(\ref{eq:xiinf}) to
be $AS_1$.
Since, in the large $|\omega|$ limit, the region $R_1$ extends to the
origin, the origin
$z=0$ and the turning point $z_T$ are connected via $AS_1$ in that limit. 
  
We now start using the WKB approximation to solve
Eq.~(\ref{eq:schrodinger}). Along $ AS_1 $, we can express $ \Psi(\xi)
$ using the standard WKB approximation:
\begin{equation}\label{eq:WKB1}
\Psi_B(\xi)=\frac{B_{+}}{\sqrt{Q}}e^{i\int^{\xi}_{\xi_\infty} d\xi' Q}+\frac{B_{-}}{\sqrt{Q}}e^{-i\int^{\xi}_{\xi_      \infty} d\xi' Q}\approx \frac{B_{+}}{\sqrt{Q}}e^{i\omega(\xi-\xi_\infty)}+\frac{B_{-}}{\sqrt{Q}}e^{-i\omega(\xi-\xi_\infty)}
\end{equation}
as long as $\xi$ stays in region $R_1$. For convenience, we choose $\xi_\infty$ 
to be the lower limit of the ``phase integration" in
Eq.~(\ref{eq:WKB1}). A different choice of the lower limit of the
integration would not affect our final results, but would complicate
their derivation. 
Matching Eq.~(\ref{eq:WKB1}) with Eq.~(\ref{eq:psia}) in region $ R_2 $, we have:
\begin{equation}\label{eq:aa}
\frac{A_{+}}{A_{-}}=\frac{B_{+}}{B_{-}}e^{-2i(\omega\xi_\infty-\delta_0)}.
\end{equation}
Similarly, along another anti-Stokes line $AS_2$ in region $R_1$, we could write $ \Psi(\xi) $ as:
\begin{equation}\label{eq:WKB2}
\Psi_C(\xi)\approx\frac{C_{+}}{\sqrt{Q}}e^{i\omega(\xi-\xi_{\infty})}+\frac{C_{-}}{\sqrt{Q}}e^{-i\omega(\xi-\xi_{\infty})}.
\end{equation}
As both $\Psi_B(\xi)$ and $\Psi_C(\xi)$ represent the same solution of
the Schr{\"o}dinger equation~(\ref{eq:schrodinger}) in different Stokes
domains, $ C_{\pm} $ can be expressed as a linear combination of
$B_{\pm} $. To see that, we trace $\Psi_B(\xi)$ along a path
$\Gamma$  connecting $AS_1$ and $AS_2$ while staying in $R_1$ as illustrated in
Fig.~\ref{fig:stokes}. Along $\Gamma$, the ``in-falling wave'' term
$e^{i\omega(\xi-\xi_{\infty})}$ is (exponentially) dominant over the
second term  $e^{-i\omega(\xi-\xi_{\infty})}$. Therefore $B_{+}$ must
not change along $\Gamma$, no matter the value of $B_-$:
\begin{equation}
C_{+}=B_{+}.
\end{equation}
On the other hand, if $B_+=0$, then $B_{-}$ cannot change along $\Gamma$, i.e., $C_{-}=B_{-}$ if $B_{+}=0$. As a result, we can express $C_{-}$ as:
\begin{equation}\label{eq:connectbc}
C_{-}=B_{-}+ S B_{+}
\end{equation}
where the multiplier $ S $ is the Stokes constant~\cite{headingbook}
with respect to point $ \xi_\infty $. The phenomenon that
the coefficient of the subdominant solution is shifted by a product of
$S$ and the coefficient of the (unchanged) dominant term is the
well-known Stokes phenomenon.\footnote{This shift occurs along $\Gamma$ discontinuously at the
crossing of the Stokes line separating the Stokes domains.}

In addition, near $z_{H}$, we have from Eq.~(\ref{eq:xi}): $ \xi\approx -\log(z_H-z)/4$. Then selecting the solution $ V_{-}(z) $ in Eq.~(\ref{eq:green}) is equivalent to imposing the infalling wave condition:
\begin{equation}\label{eq:bhs}
C_{-}=0
\end{equation}  
along $ AS_2 $~\cite{Natario:2004jd}. From Eq.~(\ref{eq:connectbc}), we obtain: 
\begin{equation}\label{eq:connecting}
\frac{B_{+}}{B_{-}}=-\frac{1}{S}=e^{-\log S-i\pi}.
\end{equation}
Substituting the above equation~(\ref{eq:connecting}) in Eq.~(\ref{eq:aa}) and using the definition~(\ref{eq:D}), we establish an asymptotic expression for $\mathcal{D}(\omega)$:
\begin{equation}\label{eq:asymd}
\omega \mathcal{D}(\omega)=\omega \xi_\infty -\frac{\pi}{4}(2\nu_0
-1)-\frac{i}{2}\log S
\end{equation} 
and from Eq.~(\ref{eq:qnmd}) the asymptotic behavior of $\omega_n$ and $ r_n$:
\begin{equation}\label{eq:wkbd}
\omega_n =\left[n +\frac{i}{2\pi}\log S+\frac{1}{4}(2\nu_0 -1)\right]{\omega_0},\qquad r_n=K\left({\omega_0}+\frac{i\omega^2_0}{2\pi S}\frac{\partial S}{\partial\omega}\Big|_{\omega=\omega_n}\right)
\end{equation}
where
\begin{equation}
  \label{eq:omega0}
  \omega_0 = \pi/\xi_\infty.
\end{equation}

\subsection{The Stokes constant}
\label{sec:stokes-constant}


To gain insight into how and whether $S$ should (or should not) depend
on $\omega$, it is useful to think of the Stokes phenomenon in the
following way~\cite{meyer1989}.  Both WKB solutions Eqs.~(\ref{eq:WKB1})
and~(\ref{eq:WKB2}), describe the {\em same} exact solution of the
Schr{\"o}dinger equation in two different Stokes sectors around the
turning point. These approximate solutions are multivalued functions
with a branching singularity at the turning point, $Q(\xi_T)=0$. However, the
exact solution of the Schr\"odinger equation is analytic at a
regular point, such as the turning point. In order to match the
absense of the branching singularity in the exact solution, the WKB
solutions must compensate their discontinuity along a path winding
around the turning point by a corresponding discontinuity in the
coefficients $B$ and $C$. This is the essense of the Stokes
phenomenon. For an {\em isolated} regular turning point this argument leads
to the well-known value of $S=i$.

More importantly, this argument sheds light on the reason why the
Stokes constant should have a different value in a special case when
the turning point approaches a singularity of the Schr{\"o}dinger
potential in the limit $|\omega|\to\infty$, as it does in our
case. Since, in this case, winding around the turning point, while
staying in $R_1$, requires winding around the singularity also (as
well as other turning points). If the singularity is
a branching point of the {\em exact} solution, the discontinuity
across the cut is reflected in the value of the Stokes constant, which
thus depends on the nature of the singularity.

In fact, since we assume that, at large $z$, $U\sim z^m$ and $z/h$
vanishes to guarantee the convergence of the integration in
Eq.~(\ref{eq:xiinf}), $\xi_{\infty}$ will always be a singular point
of Eq.~(\ref{eq:schrodinger}).  If $\xi_\infty$ is a {\em regular}
singular point of Eq.~(\ref{eq:schrodinger}), $\Psi(\xi)$ can be
expressed as a linear combination of two Frobenius series solutions:
$(\xi-\xi_{\infty})^{f_{\pm}}(1+O(\xi-\xi_{\infty}))$ with $f_{\pm}$ being
the indicial exponents and $f_{+}+f_{-}=1$.  Taking the WKB solution
around the point $\xi_\infty$ and matching the discontinuity of the
exact solution, one finds the
Stokes constant: $S=2i\cos[\pi(f_{+}-f_{-})/2]$~\cite{meyer1989} as we
explain in detail in Appendix~\ref{sec:stokes}. If the
 indicial exponents $f_{\pm}$ are independent of $\omega$,
the resulting Stokes constant has no $\omega$ dependence either.

Even if $\xi_{\infty}$ is an {\em irregular\/} singular point of
Eq.~(\ref{eq:schrodinger}) one may still expect that determining $S$,
though more involved, is still possible, perhaps along the lines of
Ref.~\cite{Meyer1983145, 10.1137/0514038} (see also
Appendix~\ref{sec:stokes}). From a more practical point of view,
which we take in Sec.~\ref{sec:softwall},
even if one has not found an easy way to determine $S$ in that
situation, one could attempt to fit asymptotic behavior of $\omega_n$
numerically using Eq.~(\ref{eq:offset}). If the quality of the fit is good
and the resulting $\omega_0$ is close to the analytical expectation
given by Eq.~(\ref{eq:xiinf}), then it is very likely that $S$ will
approach a constant in large $|\omega|$ limit. In fact, that is what
we observe for the soft-wall model at finite temperature (see
Sec.~(\ref{sec:softwall}) below).

In conclusion, we anticipate, on general grounds, that for a large
class of theories the Stokes constant $S$ is a finite constant in the
large $|\omega|$ limit. As a result, the asymptotics~(\ref{eq:gapoff})
are established.

Furthermore, to show that the summation over $(r_n-K\omega_0)$ is
convergent in the sum rule~(\ref{eq:srs}), we need to show that
summation over $S^{-1}\partial S/\partial\omega$ terms is convergent. For
sufficiently large $n$, we can replace the summation with
integration. Then the {\em existence} of the large $|\omega|$ limit of $\log S$
would imply that the integration of $S^{-1}\partial
S/\partial\omega$ is convergent and complete the derivation of the conductivity sum rule~(\ref{eq:srs}).

\subsection{Examples and comparisons}
\label{sec:examples}

The authors of Ref.~\cite{Natario:2004jd} have considered the cases that the Schr{\"o}dinger Equation~(\ref{eq:schrodinger}) takes the form:
\begin{equation}\label{eq:nearinfty}
\frac{d^2\Psi(\xi)}{d\xi^2}+\left[\omega^2 -\frac{\nu^2_{\infty}-\frac{1}{4}}{(\xi-\xi_\infty)^2}\right]\Psi(\xi)=0,\qquad\text{when}\qquad |z|\to\infty.
\end{equation}
Then $f_{\pm}=\pm\nu_{\infty}-1/2$ and we have $S=2i\cos(\pi\nu_{\infty})$(see also Ref.~\cite{headingbook}). Consequently, we read from Eq.~(\ref{eq:wkbd}) that:
\begin{equation}\label{eq:offset}
\Delta =\left[\frac{i}{2\pi}\log(2i\cos(\pi\nu_{\infty} ))+\frac{1}{4}(2\nu_{0}-1)\right]{\omega_0},
\end{equation} 
in complete agreement of the results of Ref.~\cite{Natario:2004jd} obtained by using the properties of the Bessel functions\footnote{We have converted the results of Ref.~\cite{Natario:2004jd} into the notations used in this paper.}. For that reason, the first part of Eq.~(\ref{eq:wkbd}) is a generalization of the previous work. Although asymptotic behavior of $ r_n $ can be calculated straightforwardly from the WKB approximations, the expression of $ r_n $ in second part of Eq.~(\ref{eq:wkbd}), to the best of our knowledge, is \textit{new}. 

Finally, let us check the results obtained in this section in the case of
pure thermal AdS black-hole background. In that case, $B(z)=-\log z$,
$U(z)\approx -5z^6/4$ and $\xi\approx\xi_\infty+1/(3z^3)$
when $|z|$ is large. In that limit, the Schr{\"o}dinger equation~(\ref{eq:schrodinger}) is reduced to Eq.~(\ref{eq:nearinfty}) with
$\nu_{\infty}=-1/3$. From Eq.~(\ref{eq:offset}) and recalling that
$\nu_0=1$, we obtain $\Delta=0$ due to the cancellation between two
terms in Eq.~(\ref{eq:offset}).  Moreover, if $h(z)$ is a polynomial,
as it is in the case at hand,  $h(z)=1-z^4$,  there is a simple way to evaluate $\xi_\infty$ in Eq.~(\ref{eq:xiinf}): 
\begin{equation}\label{eq:xiinfc}
\xi_{\infty}=\frac{1}{2}\int^{z_\infty}_{-z_\infty}dz\frac{1}{h(z)}=\frac{1}{2}\oint_{\mathcal{C}}dz\frac{1}{h(z)}=-\pi i\sum_{z_h}\frac{1}{h'(z_h)}.
\end{equation} 
In the first equality we have used the property $h(z)=h(-z)$. The
contour $\mathcal{C}$ is chosen to connect $\pm z_{\infty}$ by a
straight line and a large semi-circle centered at the origin of the
complex $z$ plane. As $|z_{\infty}|\to \infty$ when
$|\omega|\to\infty$, the contribution from the integration along the
semi-circle vanishes for $h(z)$ being a polynomial of $z^2$. Applying
the Cauchy integral theorem to the integral, we obtain the rightmost
expression in Eq.~(\ref{eq:xiinfc}). The summation here denotes the
summation over all $z_h$s, the zeros of $h(z)$, enclosed in the
contour $\mathcal{C}$. In particular, for the pure AdS black-hole
background, ${\rm Arg\,}z_\infty=\pi/12$ as can be seen in
Fig.~\ref{fig:inz}. Therefore $z=1,-i$ are the zeros of $h(z)$
enclosed in the contour $\mathcal{C}$. Consequently,
$\xi_\infty=(1+i)/T$ thus $\omega_0=\pi/\xi_{\infty}=2(1-i)\pi T$ (we
restored the units which were set by $\pi T=1$ in this Section). One
can check that $\omega_n, r_n$ given by Eq.~(\ref{eq:gapoff}) coincide
with the quasi-normal spectrum of $\mathcal{N}=4$ SYM theory given by
Eq.~(\ref{eq:qnma}).

\section{\label{sec:softwall} Examining the sum rule in the``soft-wall" model at finite temperature.}

In this section, we will examine the sum rule~(\ref{eq:srs}) with the ``soft-wall"
model~\cite{Karch:2006pv}, a holographic QCD model, at finite
temperature. With $\phi(z)=cz^2$ and $A(z)=-\log z$, the ``soft-wall"
model reproduces the Regge-like trajectory of the vector mesons,
$m^2_n=4nc$, at zero temperature~\cite{Karch:2006pv}. Studying that
model at finite temperature can provide a non-trivial check of the sum
rule~(\ref{eq:srs}). It can also illustrate how the dissociation or
``melting'' of the bound-states is related to the increase in
conductivity, the phenomenon which is relevant to the transport
properties of sQGP.

Dimensionless ratios of physical quantities in the ``soft-wall'' model
at finite temperature are fully controlled by the dimensionless
parameter $\tilde{c}=c/(\pi T)^2$. To make the connection with the
real world more tangible, we set the overall scale by taking $c=2.54$
GeV to fit the mass of the $J/\psi$ at zero temperature. Such a choice
has been used in Ref.~\cite{Fujita:2009ca, Fujita:2009wc} to study
the thermal charmonium spectral functions.\footnote{With this choice,
  $J/\psi$ decay constant and the $\psi'$ mass and decay constant are
  off by nearly $20\%$. This is because while there is only one
  parameter $c$ in the ``soft-wall" model, the spectrum of the
  quarkonium at zero temperature is controlled by both heavy quark
  masses and the string tension (or $\Lambda_{QCD}$). A more realistic
  holographic model of charmonium addressing this issue can be found
  in Ref.~\cite{Grigoryan:2010pj}.}

\begin{figure}
\centering
\includegraphics{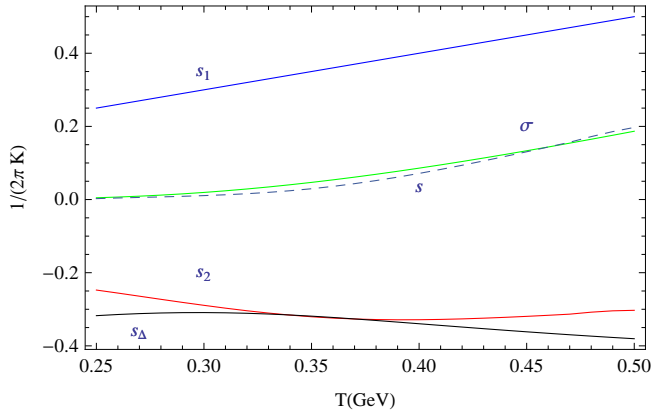} 
    \caption{\label{fig:sw}  $\sigma$(Green), $s$(dashed),
      $s_1$(blue), and $s_2$(red) divided by the
      constant $2\pi K$ as a function of $T$. To evaluate $s_2$, we
      truncated the summation at $n=5$. We also plot the contribution
      from the $\Delta$ term to the total sum normalized by $2\pi K$: $s_\Delta=-2K{\rm Im\,}\Delta$ in black.} 
\end{figure}

Following~Ref.~\cite{Fujita:2009ca,    Fujita:2009wc,Grigoryan:2010pj,Herzog:2006ra}), we assume pure
black-hole metric background, i.e, $h(z)=1-(z\pi T)^4$. We have
calculated the first five quasi-normal modes $\omega_n$ and the
corresponding rescaled residues $r_n$ numerically for temperature $T$
between $250$ MeV and $500$ MeV. For the soft-wall model at finite
temperature, the point $\xi_{\infty}$ is an irregular singular
point of Eq.~(\ref{eq:schrodinger}). As we have explained in the previous
section, we fit $\omega_n$ using Eq.~(\ref{eq:offset}) to
obtain $\Delta$. Indeed, Eq.~(\ref{eq:offset}) provides a good fit for
$\omega_n$s where $n=2,3,4,5$ and the resulting $\omega_0$ is close to
the expected asymptotic value $2(1-i)\pi T$. To analyze separate
contributions to the conductivity, we split the R.H.S of the
sum rule~(\ref{eq:srs}) into two terms:
\begin{equation}\label{eq:split}
s_1=-{K}{\rm Im\,}(\omega_0),\qquad\text{and}\qquad
s_2=-{2K}{\rm Im\,}\Delta+2\sum^{n_{\rm max}}_{n=1}{\rm Im\,}\left(r_n-K{\omega_0}\right).
\end{equation}
where in practice we set $n_{\rm max}=5$.
 In Fig.~\ref{fig:sw}, we plot $\sigma,s, s_1,s_2$ and the total sum
 $s=s_1+s_2$ (normalized by $2\pi K$) versus $T$. We extract the conductivity via the analytic results of Ref.~\cite{Iqbal:2008by}(see also Ref.~\cite{Grigoryan:2010pj}):
\begin{equation}\label{eq:conducivitya}
\sigma= 2\pi KT e^{B(z_H)}.
\end{equation}
Obviously from the plot, the R.H.S of the sum rule~(\ref{eq:srs}) $s$
(dashed line), calculated numerically, is close to the conductivity
$\sigma$(solid green line) given by Eq.~(\ref{eq:conducivitya}) in the
temperature range we are considering. We take that result as a
numerical evidence that the sum rule~(\ref{eq:srs}) applies to the
``soft-wall" model. Moreover, $s_1$ is linear in $T$ and has no $c$
dependence. Thus, we could interpret $s_1$ as the contribution to the
conductivity from the thermal AdS background with no account of
confinement effect introduced by parameter $c$. We also observe that
$s_2$ is always negative. That can be thought of as a reflection of
the physical fact that the presence of bound states reduces the number
of the charge carriers in medium and lowers the
conductivity. That effect is quantified by $s_2$.  Finally, we note
from Fig.~\ref{fig:sw} that $s_2$ in Eq.~(\ref{eq:split}) is
dominated by $s_\Delta=-2K{\rm Im\,}\Delta$ term in the range of the
temperature we are studying. That would mean that in some cases, one
may be able to use $-K{\rm Im\,}(\omega_0+2\Delta)$ as a reasonable
estimate of $\sigma$.

\section{\label{sec:conclusion}Summary and discussion}

We have shown that the current-current correlator in a theory with
holographic dual description can be represented as a convergent infinite
sum over the quasinormal mode poles Eq.~(\ref{eq:grep}). We have established the
convergence by deriving the asymtpotic behavior of the quasinormal
mode frequencies and residues Eq.~(\ref{eq:wkbd}) 
using the WKB, or phase integral, approach. 

We have also established a sum rule relating conductivity $\sigma$ to
the convergent infinite sum over quasinormal modes
Eq.~(\ref{eq:srs}). We have checked this sum rule in the exactly
solvable case of the ${\cal N}=4$ SUSY Yang-Mills theory.  We studied the
non-trivial example of the soft-wall holographic model numerically and
found that the sum rule is in good agreement with analytically known
value of the conductivity, and that the sum over the quasinormal modes
is quickly saturated by a few lowest terms.

\subsection{Spectral function}
\label{sec:spectral-function}

Using representation Eq.~(\ref{eq:grep}) for $G_R$ we can also obtain a
corresponding convergent representation for the spectral function:
\begin{equation}\label{eq:greprho} 
\rho(\omega)=\sigma\omega-\omega^2 \sum_{n}{\rm Im\,}\left[\frac{r_n}{\omega-\omega_n}+\frac{\tilde{r}_n}{\omega-\tilde{\omega}_n}\right].
\end{equation}

We have expressed the ``gap'' $\omega_0$ and ``offset'' $\Delta$
parameters of the quasinormal modes in terms of the singular point of
the Schr{\"o}dinger quation $\xi_\infty$ and the corresponding
Stokes constant $S$, Eq.~(\ref{eq:wkbd}),~(\ref{eq:omega0}). 
Further insight into the significance of $\xi_{\infty}$ (or $\omega_0$) and
$S$ (or $\Delta$) may be obtained if one assumes that asymptotic
expression of $ \mathcal{D}(\omega) $ in Eq.~(\ref{eq:asymd}) can be
continued to the real axis ${\rm Arg}(\omega)=0$. Let us further
assume that $U(z)$ and $h(z)$ are even functions of $z$. Then $\xi(z)$
defined by Eq.~(\ref{eq:xi}) is an odd function of $z$. Consequently,
$U(\xi)$ is an even function of $\xi$. One can then argue that the corrections to
Eq.~(\ref{eq:wkbg}) should be in even powers of $\omega^{-1}$. However, as
$\rho(\omega)$ is an odd function of $\omega$, those power corrections
may not affect the asymptotic behavior of $\rho(\omega)$\footnote{This
  is in agreement with the results of Ref.~\cite{CaronHuot:2009ns}
  that spectral densities have no power law corrections in asymptotic
  expansion if the OPE of Euclidean correlators are free from
  non-analytic terms.}. We could then use Eq.~(\ref{eq:wkbg}) and
Eq.~(\ref{eq:asymd}) to study the asymptotics of the spectral density:
\begin{equation}\label{eq:asyrho}
\rho(\omega)\to\pi K\omega^2\left[1+2{\rm Im\,}(Se^{2i\omega\xi_\infty})\right]=\pi K\omega^2
\left[1+2e^{-2\omega\xi_I}{\rm Im\,}(Se^{2i\omega\xi_R})\right]
\end{equation} 
where $\xi_\infty=\xi_R+i\xi_I$. The first term in the square
brackets, $ 1 $, on R.H.S of Eq.~(\ref{eq:asyrho}) is expected as $
\rho(\omega) $ will asymptotically approach zero temperature
limit. The next term explains the observation made on the basis of the
numerical studies of Ref.~\cite{Teaney:2006nc} that ``finite
temperature result oscillates around the zero temperature result with
exponentially decreasing amplitude." The author of
Ref.~\cite{Teaney:2006nc}  argues that such behavior is intimately
connected with the analytic structure stemming from the quasi-normal
modes. Indeed, since ${\rm Im\,}\omega_0<0$ and thus
$\xi_I>0$, Eq.~(\ref{eq:asyrho}) shows that $2\xi_R $ and $ 2\xi_I$
correspond to the oscillation frequency and the damping rate
respectively. Our analysis suggests that such phenomenon is quite
generic for theories with a gravity dual.

One can easily check the correctness of Eq.~(\ref{eq:asyrho}) with
Eq.~(\ref{eq:adsg}) for the $\mathcal{N}=4$ SYM in the strong coupling
limit where the Green's function is known
analytically~\cite{Myers2007}. In addition,
one can extend our analysis to other channels, e.g., the shear
channel, as well. For example, for $\mathcal{N}=4$ SYM in the strong
coupling limit, again, $\xi_\infty=(1+i)/4T$, we then predict the
damping rate of the corresponding spectral density to be $1/(2T)$
while by fitting numerics, the authors of
Ref.~\cite{Romatschke:2009ng} obtained a damping rate of $.46/T$.

In passing, we also note that due to the asymptotic behavior in
Eq.~(\ref{eq:asyrho}), the integral over
$\omega^{2n-1}\delta\rho(\omega)$, where $\delta\rho(\omega)=\rho(\omega,T)-\rho(\omega,T=0)$, is convergent for any positive integer
$n$. This suggests that for theories with a gravity dual, one could
establish a family of f-sum rules~\cite{forster1975hydrodynamic} as
discussed in Appendix~\ref{sec:fsum}.

\subsection{Conductivity}
\label{sec:conductivity}

An insight into the meaning of the conductivity sum rule can be obtained
by assuming a naive representation of the Green's function $G_R(\omega)$ in terms of
the quasinormal modes:
\begin{equation}
  \label{eq:GR-naive}
   G_R(\omega) \stackrel{R}{=} \sum_n
\left[\frac{\omega_n^2 r_n}{\omega-\omega_n}+\frac{\tilde{\omega}^2_n\tilde r_n}{\omega-\tilde\omega_n}\right]
\end{equation}
This representation ignores the fact the the sum is divergent. In a
certain sense, the convergent representation~(\ref{eq:grep}) is a
regularized version of the naive representation (which is indicated by
letter $R$ in~(\ref{eq:GR-naive})). Taking imaginary part in
Eq.~(\ref{eq:GR-naive}) and using Kubo formula~(\ref{eq:conductivity})
we would find
\begin{equation}
  \label{eq:sigma-naive}
  \sigma \stackrel{R}{=} 2{\rm Im\,} \sum_n r_n.
\end{equation}
Again, this sum rule ignores divergence of the
sum. We can think of Eq.~(\ref{eq:srs}) as the regularized
form of this naive sum rule.

One could interpret the naive sum rule~(\ref{eq:sigma-naive}) as an
expression of the following physical picture. Consider the behavior of
quarkonia-like resonances as a function of temperature. As the
temperature is increased the resonance poles in the Green's function
move into (the lower half of) the complex plane. The residues $r_n$,
starting off as the real decay constants at $T=0$, acquire their
imaginary parts at finite temperature and are thus related to the
process of ``melting'' or dissociation of the resonances. As a bound
state dissociates, the conductivity receives contribution from the
freed charge carriers.

Although this picture is intuitive, its usefulness is limited, as the
actual example we considered in Section~ \ref{sec:softwall} shows. We
find that the dominant contribution to the conductivity comes from the
first term $-K{\rm Im\,}(\omega_0+2\Delta)$ of the sum rule
Eq.~(\ref{eq:srs}). This term could be thought of as the combined contribution
of the whole tower of resonances as it is a result of the regularization of
the divergent sum in Eq.~(\ref{eq:sigma-naive}).

\acknowledgments
We would like to thank the Institute for Nuclear Theory at
the University of Washington for hospitality during the INT
Summer School on Applications of String Theory, when part of this
work was carried out. We thank Todd Springer for reading the draft and
suggestions. Y.Y. would like to thank Yang Zhang for discussions.
This work is supported by the DOE grant No.\ DE-FG0201ER41195. 

\appendix

\section{\label{sec:redundant} Redundant poles and  $\omega=-2in\pi T$ }
We now discuss a subtle issue in the context of gauge/gravity dual on how to define $ G_R(\omega) $ at following points:
\begin{equation}\label{eq:redundant}
\omega= -2n i\pi T, \qquad n=1,2,\ldots.
\end{equation}
We consider the Frobenius power series expansion Eq.~(\ref{eq:Frobenius}) near $ z_H $:
\begin{equation}\label{eq:Frobenius}
V_{\pm}(z,\omega)=(z-z_H)^{r_{\pm}}\sum^{\infty}_{j=0}c^{\pm}_{j}(\omega)(z-z_H)^j
\end{equation}
where the indicial exponents $ r_{\pm}=\pm i\omega/4\pi T $ and~\cite{coddingtonbook}:
\begin{equation}\label{eq:cminus}
c^{-}_{j}(\omega)=\frac{F^{-}_{j}}{j(i\omega/(2\pi T) -j)}.
\end{equation}
Here, $F^{-}_{j}$ is a linear combination of $ c^{-}_{j-1},\ldots,c^{-}_{0}$~\cite{coddingtonbook}:
\begin{equation}\label{eq:F}
F^{-}_{j}=\sum^{j-1}_{k=0}[(k+r_{-})\alpha_{j-k}+\omega^2\beta_{j-k}]c^{-}_{k}
\end{equation} 
and $\alpha_j, \beta_j$ have no $\omega$ dependence~\cite{coddingtonbook}:
\begin{equation}\label{eq:albe}
(z-z_H)(B'+\frac{h'}{h})=\sum^{\infty}_{k=0}\alpha_{k}(z-z_H)^k,\qquad \frac{(z-z_H)^2}{h^2}=\sum^{\infty}_{k=0}\beta_{k}(z-z_H)^k.
\end{equation}
As a result of Eq.~(\ref{eq:cminus},\ref{eq:F},\ref{eq:albe}), the coefficient  $ c^{-}_{j}(\omega) $ will be a meromorphic function of $\omega$ with simple poles at $ \omega $ given by Eq.~(\ref{eq:redundant}) for $ j\ge n $.
However, those singularities can be cured naturally by suitably choosing the overall constant $ c^{-}_0 $. For example, one may define\footnote{This trick has been used to analytically continue the Gauss hypergeometric function in its parameter space.}:
\begin{equation}
c^{-}_0=\frac{1}{\Gamma(1-i\omega/(2\pi T))},
\end{equation} 
then the Frobenius solutions~(\ref{eq:Frobenius}) are regular in the entire complex $ \omega$ plane. Consequently, those points listed by Eq.~(\ref{eq:redundant}) will not, in general, lead to any additional singularities of $G_R(\omega)$. Noting from Eq.~(\ref{eq:green}) that $ G_R(\omega) $ has no dependence on the overall normalization of the solution $ V_{-}(\omega,z) $, a different choice of $c^{-}_{0}$ will not affect resulting $G_R(\omega)$. 

In fact, those special points have been known as ``redundant
zeros (poles)"~\cite{PhysRev.69.668} since long time ago in the context of the non-relativistic scattering. It has been shown~\cite{Proc.Roy.Soc, J.Math.Physic.1} that for the Schr{\"o}dinger potential with the exponential tail:
\begin{equation}
U(\xi)\sim e^{-y\xi}\qquad \xi\to\infty
\end{equation}
the infalling wave solutions $ \Psi(\xi,\omega) $ will have simple poles at $ \omega=-in y/2$. 
One can check from Eq.~(\ref{eq:schpotential}) and Eq.~(\ref{eq:xi}) that in our case, $ y=4\pi T $, thus again we have Eq.~(\ref{eq:redundant}).
Historically, those poles are called ``redundant poles'' because they do not represent the true resonant states. 
In the context of the holographic correspondence, the name ``redundant poles" may still be appropriate as they are not related to the singular points of the retarded Greens function $ G_R(\omega) $. As a result, the singularities of $ G_R(\omega) $ are only due to simple poles at $ \omega_n(\tilde{\omega}_n) $.

\section{\label{sec:stokes} The Stokes constant for a regular singular point}
In this section, we will derive the Stokes constant with respect to a regular singular point of a Schr{\"o}dinger-type equation
\begin{equation}\label{eq:sch}
\frac{d^2\Psi}{dy^2}+\left(\lambda^2 -U_0(y)\right)\Psi(y)=0.
\end{equation} 
Without losing generality, we set $y=0$ to be that regular singular point such that:
\begin{equation}
\mathop{\lim}\limits_{y\to 0}y^2U_0(y)=l^2
\end{equation}
and $\lambda$ to be a real positive large parameter.  Eq.~(\ref{eq:sch}) has two Frobenius series solutions: 
\begin{subequations}\label{eq:fro}
\begin{equation}
\psi_1(y)=y^{f_{+}}g_1(y)
\end{equation}

\begin{equation}
\psi_2(y)=y^{f_{-}}g_2(y)
\end{equation}
\end{subequations}
where $g_1(y),g_2(y)$ are power series of $y$. The indicial exponents are the roots of the equation:
\begin{equation}
f^2-f-l^2=0.
\end{equation}
One may note from Weda's theorem that
\begin{equation}
f_{+}+f_{-}=1.
\end{equation}
A solution of Eq.~(\ref{eq:sch}), $\Psi(y)$,  can be expressed as a linear combination of $\Psi_1(y),\Psi_2(y)$, i.e.,
\begin{equation}
\Psi(y)=c_1\Psi_1(y)+c_2\Psi_2(y).
\end{equation}
From Eq.~(\ref{eq:fro}), we also have:
\begin{subequations}
\begin{equation}\label{eq:cclock}
\Psi(ye^{2\pi i})=c_1e^{2\pi if_{+}}\Psi_1(y)+c_2e^{2\pi if_{-}}\Psi_2(y);
\end{equation}
\begin{equation}\label{eq:clock}
\Psi(ye^{-2\pi i})=c_1e^{-2\pi if_{+}}\Psi_1(y)+c_2e^{-2\pi if_{-}}\Psi_2(y).
\end{equation}
\end{subequations}
Multiplying Eq.~(\ref{eq:cclock}) and Eq.~(\ref{eq:clock}) by $e^{-\pi i(f_++f_-)}$ and $e^{\pi i(f_++f_-)}$ respectively, then adding the results together, we have a connecting relation~\cite{olver1974asymptotics}: 
\begin{equation}\label{eq:connect1}
2\cos[\pi(f_{+}-f_{-})]\Psi(y)=e^{-\pi i(f_{+}+f_{-})}\Psi(y e^{2\pi i})+e^{\pi i(f_{+}+f_{-})}\Psi(y e^{-2\pi i})=-[\Psi(y e^{2\pi i})+\Psi(y e^{-2\pi i})].
\end{equation}
If we solve the Schr{\"o}dinger equation~(\ref{eq:sch}) using WKB
approximation, we find two turning points at $y=\pm l/\lambda$. For
$\lambda\to\infty$ these points approach the singularity at $y=0$. In
the region of $y$ not very close to the origin, where
$\lambda^2\gg| U(y)|$, the solution $\Psi(y)$ can be approximated by a
linear combination of two WKB solutions: $e^{\pm i\lambda y}$.

\begin{figure}
\centering
\includegraphics{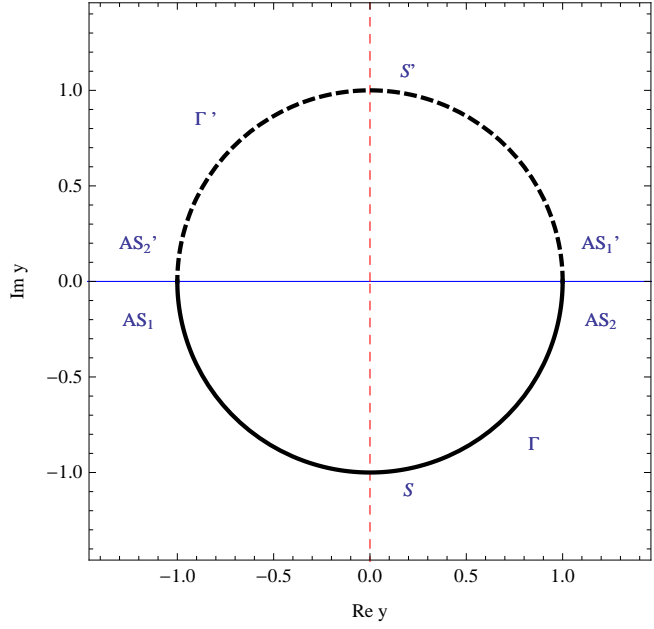} 
    \caption{\label{fig:s} The schematic plot of the Stokes diagram for Eq.~(\ref{eq:sch}). One can compare it with Fig.~\ref{fig:inxi}. In large $|\omega|$, two turning points $\xi_T,\xi_T'$ are shrunk to $\xi_\infty$. What is more, the distance between $AS_1$ and $AS'_2$(or $AS_2$ and $AS'_1$) also vanishes. Therefore, schematically, we only draw two anti-Stokes lines in blue and two Stokes lines in red dashed line in this figure. We also plot the path $\Gamma$ in black thick line and $\Gamma'$ black thick dashed line. } 
\end{figure}

We plot the Stokes diagram schematically in Fig.~\ref{fig:s}. For
large $\lambda$ the region where the WKB approximation breaks down
shrinks to the origin. This region includes both turning points and
the singularity at $y=0$. Since we are working outside that region,
so that $\lambda^2\gg| U(y)|$, we can represent this non-WKB region by a
single point at $y=0$. Only four anti-Stokes lines emanate from this
region as shown in Fig.~\ref{fig:s}: two from each turning point,
following the real axis in positive and negative directions.

To determine the Stokes constant, we will consider a WKB solution defined by its value along the real axis:
\begin{equation}
\Psi_0(y)=e^{i\lambda y},\qquad\text{when Arg$\,y=0$}.
\end{equation}

When continued \textit{counterclockwise} from the positive real axis to the negative axis along $\Gamma'$, $\Psi_0$ is unchanged as $e^{i\lambda y}$ is subdominant compared to $e^{-i\lambda y}$ in the upper half plane. If we go on continuing $\Psi_0(y)$ along $\Gamma$ from the negative real axis to the positive real axis, we will have:
\begin{equation}\label{eq:ccwkb}
\Psi_0(ye^{2\pi i})=e^{i\lambda y}+Se^{-i\lambda y}.
\end{equation} 
due to the Stokes phenomenon. Similar, when $\Psi_{0}(y)$ is continued \textit{clockwise} from the positive real axis to the negative real axis along $\Gamma$, we have:
\begin{equation}
\Psi_{0}(ye^{-\pi i})=e^{i\lambda y}-Se^{-i\lambda y}.
\end{equation}
Furthermore, when $\Psi_{0}(ye^{-\pi i})$ is continued \textit{clockwise} from the negative real axis to the positive real axis along $\Gamma'$, we obtain:
\begin{equation}\label{eq:cwkb}
\Psi_{0}(ye^{-2\pi i})=(1+S^2)e^{i\lambda y}-Se^{-i\lambda y}.
\end{equation}
Substituting Eq.~(\ref{eq:ccwkb}) and Eq.~(\ref{eq:cwkb}) in Eq.~(\ref{eq:connect1}) and comparing the coefficient of $e^{\pm i\lambda y}$, we have:
\begin{equation}
S=\pm 2i\cos\left[\frac{\pi(f_{+}-f_{-})}{2}\right],
\end{equation} 
the desired Stokes constant. 

As pointed out in Ref.~\cite{olver1974asymptotics}, if $y$ is an irregular singular point, in general, it is still true that there exist solutions with the properties:
\begin{subequations}\label{eq:irre}
\begin{equation}
\psi_1(ye^{2i\pi})=e^{2\pi if_{+}}\psi_1(y)
\end{equation}
\begin{equation}
\psi_2(ye^{2i\pi})=e^{2\pi if_{-}}\psi_2(y)
\end{equation}
\end{subequations}
with $f_{\pm}$ called ``circuit exponents"~\cite{olver1974asymptotics} of the singularity, in analogy with indicial exponents. One then observes immediately the connecting relation~(\ref{eq:connect1}) is also true for $y$ being an irregular singular point. Because of that, one may generalize the present approach to determine $S$ to the cases that $y$ is irregular as well. 

\section{\label{sec:fsum} f-sum rules from holography}
We now derive a family of f-sum rules ~\cite{forster1975hydrodynamic} for theories with a gravity dual. Due to the representation~(\ref{eq:grep}), we have the following dispersion relation:
\begin{equation}\label{eq:dispersion}
\delta G_R(i\omega_E)=\mathcal{P}(\omega)+\int^{\infty}_{-\infty}\frac{d\omega}{2\pi}\frac{\delta\rho(\omega)}{\omega-i\omega_E}
\end{equation} 
where $\mathcal{P}(\omega)$ is a polynomial of $\omega$. We denote by
``$\delta$" the difference between the value of a function at
temperature $T$ and the value of that function at zero
temperature. For example, $\delta
G_R(\omega,T)=G_R(\omega,T)-G_R(\omega,T=0)$.  Applying the Borel
transformation to Eq.~(\ref{eq:dispersion}), we have:
\begin{equation}\label{eq:fourier}
\hat{\mathcal{B}}_{1/t_B}\delta G_R(\omega_E)=-
2t_{B}\int^{\infty}_{0}\frac{d\omega}{2\pi}\delta\rho(\omega)\sin(\omega t_B)
\end{equation}
where we have used the fact that $\rho(\omega)$ is an odd function of $\omega$. We then consider the asymptotic expansion of $ \delta G_R(i\omega_E) $ 
 \begin{equation}\label{eq:OPEa}
\delta G_{R}(i\omega_E)=\sum^{\infty}_{n=0}\frac{\delta h_n}{\omega_{E}^{2n}},\qquad\text{when}\qquad\omega_E\to\infty
\end{equation}
where the coefficients $\delta h_n$ may be calculated from OPE. Applying the Borel transformation to Eq.~(\ref{eq:OPEa}), we have:
\begin{equation}\label{eq:Zt1}
\hat{\mathcal{B}}_{1/t_B}\delta G_R(\omega_E)=\sum^{\infty}_{n=0}\frac{\delta h_n}{(2n-1)!}t_{B}^{2n}.
\end{equation}
Therefore $\delta h_n $ can be extracted by comparing the Taylor expansion coefficients of Eq.~(\ref{eq:Zt1}) with that of Eq.~(\ref{eq:fourier}):
\begin{equation}\label{eq:fsumh}
\delta h_n=\mathop{\lim}\limits_{t_B\to 0}\frac{d^{2n-1}}{dt_{B}^{2n-1}}\left[\int^{\infty}_{0^{+}}\frac{d\omega}{2\pi}\delta\rho(\omega)\sin(\omega t_B)\right]=2(-1)^n\int^{\infty}_0\frac{d\omega}{2\pi}\omega^{2n-1}\delta\rho(\omega).
\end{equation}
As the integral is convergent, we have interchanged the sequence of taking the limit and the integration. In literature, Eq.~(\ref{eq:fsumh}) is related to the f-sum rule. From the definition of $\rho(t)=\langle [J^{i}(t),J^{i}(0)]\rangle$ and Heinseberg's equation of motion, we have:
\begin{equation}
\frac{d^{2n-1}}{d t^{2n-1}}\rho(t)=(-i)^{2n-1}\langle\underbrace{[\dots[}_
{2n}J^{i}(t),J^{i}(0)],\underbrace{T^{00}(t)],\dots,T^{00}(t)]}_
{2n-1}\rangle.
\end{equation}  
By the Fourier transformation and taking $t\to 0$ limit, we have:
\begin{equation}\label{eq:fsum}
\int^{\infty}_{-\infty}\frac{d\omega}{2\pi}\omega^{2n-1}\rho(\omega)=\langle\underbrace{[\dots[}_
{2n}J^{i}(0),J^{i}(0)],\underbrace{T^{00}(0)],\dots,T^{00}(0)]}_
{2n-1}\rangle.
\end{equation}
Comparing sum rules~(\ref{eq:fsum},\ref{eq:fsumh}), we have established an expression of $\delta h_n$:
\begin{equation}
\delta h_{n+1}=(-1)^{n}\delta\langle\underbrace{[\dots[}_
{2n}J^{i}(0),J^{i}(0)],\underbrace{T^{00}(0)],\dots,T^{00}(0)]}_
{2n-1}\rangle_{T}.
\end{equation} 
Finally, we list all the formulas of the Borel transformation that we have used
\begin{equation}
\hat{\mathcal{B}}_{1/t_B}\left(\frac{1}{\omega + s}\right)=t_{B}e^{-st_B},\qquad \hat{\mathcal{B}}_{1/t_B}\left(\frac{1}{\omega^{n}} \right)=\frac{1}{(n-1)!}t^{n}_B,\qquad \hat{\mathcal{B}}_{1/t_B}\left(\log\omega\right)=-1
\end{equation}
for reference. We have also used the property that $\hat{\mathcal{B}}$
gives zero when acting on polynomials of $\omega$. 

\bibliographystyle{JHEP}
\bibliography{ref}

\end{document}